\documentclass[12pt]{article}
\usepackage{graphicx}
\newcommand{\be}{\begin{equation}}
\newcommand{\ee}{\end{equation}}
\newcommand{\beqy}{\begin{eqnarray}}
\newcommand{\eeqy}{\end{eqnarray}}
\newcommand{\jpsi}{J/\psi}
\newcommand{\etac}{\eta_c}
\newcommand{\as}{\alpha_s}
\newcommand{\etab}{\eta_b}
\newcommand{\aqed}{\alpha}

\begin{document}

\title{$\etab$ Decay into Two Photons}

\author{Nicola Fabiano\\
\textsl{ Perugia University and INFN, via Pascoli I-06100, Perugia, Italy}}
\date{}

\maketitle

\begin{abstract}
We discuss the theoretical predictions for the two photon decay width of the 
pseudoscalar $\etab$ meson. Predictions from potential models are examined.
It is found that various models are in good agreement with each other.
Results for $\etab$ are also compared with those from $\Upsilon$ data 
through the NRQCD procedure. 

\end{abstract}

\section{Introduction}
The $\etab$, the lightest of the $b\overline{b}$ bound states, hasn't been
observed yet. For this meson, $J^{PC} = 0^{-+}$, thus the two photon 
collisions are
appropriate for the investigation on the state. The LEP II is particularly 
suitable for this search because of high energy, high luminosity, high 
$\gamma \gamma$ cross section and low background from other  processes.
ALEPH~\cite{ARMIN,ALEPH} has recently started to search for $\eta_b$ into two
photons. One candidate event is found in the six--charged
particle final state and none in the four--charged particle final state, giving
the upper limits:
$ \Gamma(\etab \to \gamma\gamma)\times BR(\etab \to 4 \textrm{ charged
particles}) < 48 \textrm{ eV , } $
$ \Gamma(\etab \to \gamma\gamma)\times BR(\etab \to 6 \textrm{ charged
particles}) < 132 \textrm{ eV .} $

Unlike the charmonium, the investigation of the bottomonium spectrum is 
still to be completed. This note is dedicated to examine various theoretical
predictions for the electromagnetic decay of the pseudoscalar $\etab$, and
to improve the first estimates given in~\cite{NOI2}.
Predictions of course will be affected by the error due to  the parametric
dependence of the given potential model, an error which can be quite 
large since most of the parameters have been tuned with the charmonium
system. It is interesting to see whether the potential
models can predict  for this decay a value within experimental reach.
The paper is organised as follows: in Sect.~2 we shall compare the 
two photon decay width with   the leptonic width of the $\Upsilon$. Sect.~3
is devoted to the potential model predictions for $\etab \to \gamma \gamma$, 
with potential given by~\cite{ROSNER,IGI,NOI1}.
In Sect. 4 we show the predictions for $\etab$ decay widths, using
the  procedure introduced in~\cite{BBL} for the description of 
mesons made out of two non relativistic heavy quarks, by means of the Non
Relativistic Quantum Chromodynamics--NRQCD. 
In Sect. 5 we  compare these different  determinations 
of the  $\etab \to \gamma \gamma$ decay width 
together with a result based on a recent two loop theoretical analysis of the 
charmonium decay~\cite{CZARNECKI}.

\section{Relation to the $\Upsilon$ electromagnetic width}
We start with the two photon decay width of a pseudoscalar
quark-antiquark bound state \cite{VANROYEN}
with first order QCD corrections \cite{BARBIERI}, which can be written 
as
\be
\Gamma(\etab\rightarrow \gamma\gamma)= \Gamma_B^P\left 
[ 1 + 
\frac{\alpha_s}{\pi} 
\left(\frac{\pi^2-20}{3} \right ) \right ]
 \label{eq:widsc1l} .
\ee
In eq.~(\ref{eq:widsc1l}),
 $\Gamma_B^P$ is  the Born decay width for a non-relativistic bound state
which can be calculated from  potential models.
A first theoretical estimate for this decay width can be obtained by 
comparing eq.~(\ref{eq:widsc1l})  with the expressions for the 
vector state $\Upsilon$~\cite{MACKENZIE}, i.e.
\be
\Gamma (\Upsilon\rightarrow e^+e^-)=\Gamma_B^V \left ( 1-
\frac{16}{3}\frac{\alpha_s}{\pi} \right )  .\label{eq:widvec1l}
\ee

The expressions in eqs.~(\ref{eq:widsc1l}) and~(\ref{eq:widvec1l}) can be 
used to estimate 
the radiative width of $\etab$ from the measured values of the
leptonic decay width of $\Upsilon$, if one assumes  the same value
for  the wave function at the origin $\psi(0)$, for both the 
pseudoscalar and the vector state.
Taking into account the spin--dependent forces in QCD, one obtains 
a correction to the potential  due to magnetic field correlations
given by the expression $8\as/9m_b^2 \mathbf{s}_1 \cdot \mathbf{s}_2 4 
\pi \delta^{(3)}(\mathbf{r}) +
4 \as/3m_b^2 (3\mathbf{s}_1\cdot \mathbf{r} \mathbf{s}_2 \cdot \mathbf{r} 
- \mathbf{s}_1 \cdot \mathbf{s}_2)1/r^3$ 
(see for instance \cite{CORNELL,EICHTEN}).
This in turn modifies the wavefunction at the origin with a contribute
proportional to $\as/m_b^2$, since
 $|\psi(0)|^2 = \mu  \langle V'(r) \rangle/2 \pi$. The spin singlet 
and triplet states wavefunctions at the origin differ therefore only to 
$\mathcal{O}(\alpha_s)$.

Taking the ratio of the eqs.~(\ref{eq:widsc1l}) and~(\ref{eq:widvec1l}) 
and expanding in $\as$, we obtain:
\be
\frac{\Gamma (\eta_b\rightarrow \gamma\gamma)}{\Gamma(\Upsilon
\rightarrow e^+e^-)}\approx
\frac{1}{3} \frac{(1-3.38\as/\pi)}{(1-5.34\as/\pi)} = 
\frac{1}{3}  \left [ 1+1.96  \frac{\alpha_s}{\pi} 
 + \mathcal{O}(\alpha_s^2) \right ] \\ .
\label{eq:ratio1loop}
\ee
The correction can be computed  from the two loop expression for $\as$
and the  value~\cite{PDG} $\alpha_s(M_Z) = 0.118\pm 0.003 $. 
Using the  renormalization group equation to evaluate 
$\alpha_s(Q=2m_b=10.0\mbox{ GeV})=0.178 \pm 0.007$, and the  latest 
measurement  
\be
 \Gamma_{exp}(\Upsilon \to e^+e^-) = 1.32 \pm 0.05 \textrm{ keV }
\label{eq:upsexpwid}
\ee
one obtains
\be
\Gamma(\etab\rightarrow \gamma\gamma) \pm
\Delta\Gamma(\etab\rightarrow \gamma\gamma) =489 \pm 19 \pm 2 \textrm{ eV ,}
\label{eq:singnaive}
\ee
where the first error comes from the uncertainty on the $\Upsilon$ experimental
width, the second error from $\alpha_s$~.

\begin{figure}[t]
\begin{center}
\includegraphics*[angle=0,width=0.7\textwidth]{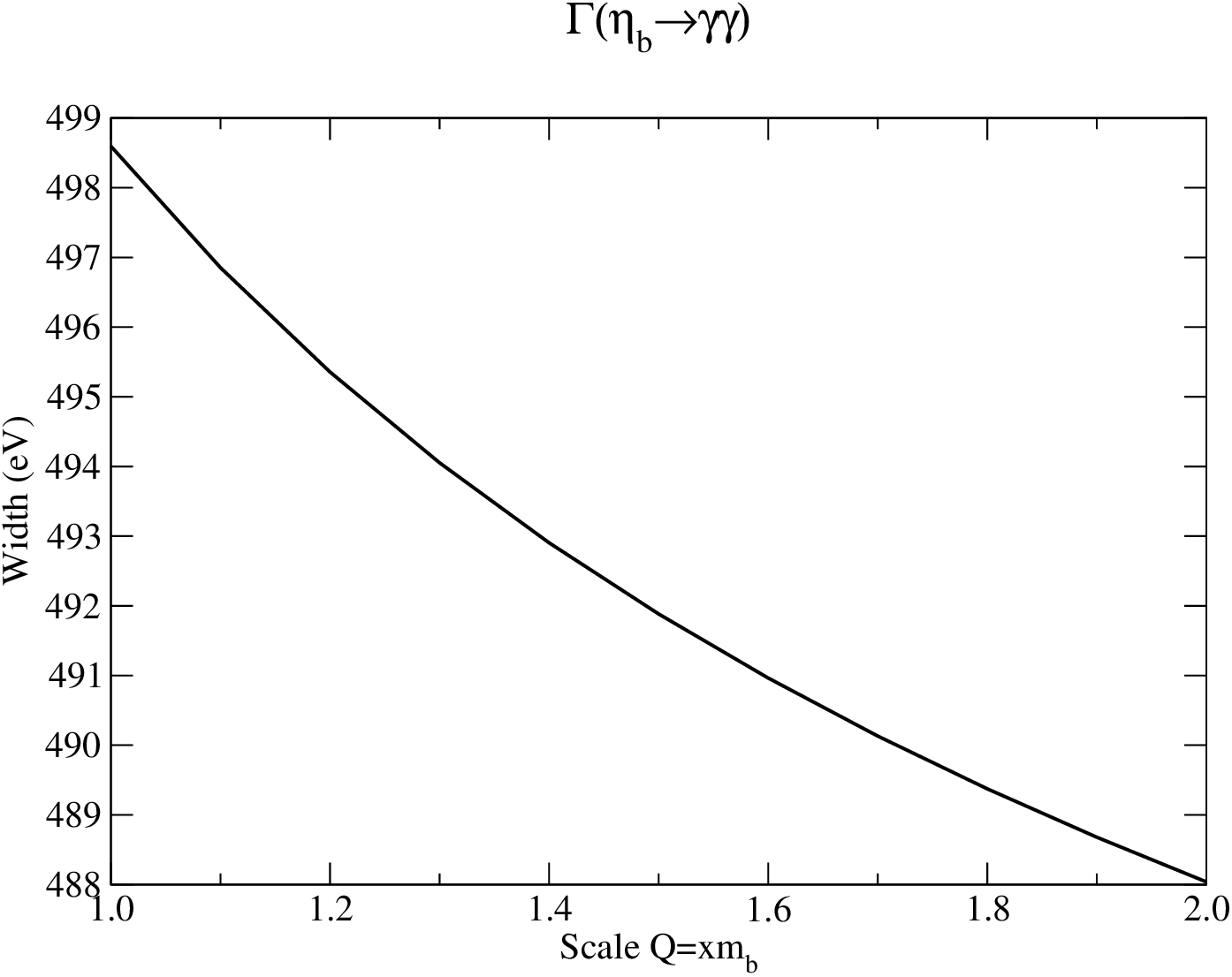}  
\caption{ The dependence  of the  $\etab$ decay width to $\gamma\gamma$ 
(in eV) is shown with respect to  the scale chosen for $\as$~in the 
radiative corrections. \label{fig:oneloop}}
\end{center}
\end{figure}

Here we have assumed the $\as$ scale to be $Q=2m_b=10.0\textrm{ GeV}$. This
choice is by no way unique as shown for the $\etac$ decay~\cite{NOI2},
 and in fig.~(\ref{fig:oneloop}) we show the 
dependence of the $\etab$ photonic  width, evaluated  from 
eq.~(\ref{eq:upsexpwid}), upon  different values of the scale 
chosen for $\as$~. Since there are no experimental measurements
of this decay we shall assume, like for the $\etac$ case, that it is not 
possible to determinate a scale choice of $\as$~.
We shall therefore include this
fluctuation in the indetermination due to radiative corrections.
\section{Potential models predictions for $\etab$}
We present now the results one can obtain for
the absolute width, through the extraction of the wave
function at the origin from potential models.
For the calculation of the wavefunction~\cite{SCHROEDINGER}
we have used four different 
potential models,
like the  potential  of Rosner et al.~\cite{ROSNER}
$ V(r) = \lambda ((r/r_0)^{\alpha} - 1 )/\alpha +C \mbox{ .}$
with $r_0 = 1 \textrm{ GeV}^{-1} , \;  \alpha=-0.14 \mbox{ , } 
\lambda=0.808\textrm{ GeV, } C=-1.305\textrm{ GeV}$, 
and  the QCD inspired potential $V_{J}$  of Igi-Ono \cite{IGI,TYE}
\be
 V_{J}(r) = V_{AR}(r) + d r e^{-gr} + ar , \ \ \
 V _{AR}(r) = -\frac{4}{3} \frac{\alpha_{s}^{(2)}(r)}{r} \label{eq:igipot}
\ee 
with two different  parameter sets, corresponding to 
 $\Lambda_{\overline{MS}}=0.5 \  GeV$
and $\Lambda_{\overline{MS}}=0.2 \  GeV$ respectively \cite{IGI}.

\begin{table}[!htb]
\begin{center}
\begin{tabular}{cccc}
$\Lambda_{\overline{MS}} (GeV)$ & $a (GeV^2)$ & $g (GeV)$ & $d (GeV^2)$ \\
\hline
0.2 & 0.1587 & 0.3436 & 0.2550  \\
0.5 & 0.1391 & 2.955 & 1.776  \\
\hline
\end{tabular}
\end{center}
\caption{ \textit{Parameters chosen for Igi--Ono potential $V_J(r)$.}}
\end{table}
We also 
show the results from a Coulombic type potential with the QCD 
coupling $\alpha_s$ frozen
to a value of $r$ which corresponds to the Bohr radius of the
quarkonium system, $r_B=3/(2m_b\as)$ (see for instance \cite{NOI1}). 
We shall stress that the scale of $\as$ occurring in the radiative correction
and the one of Coulombic potential are different.

\begin{figure}[t]
\begin{center}
\includegraphics*[angle=0,width=0.7\textwidth]{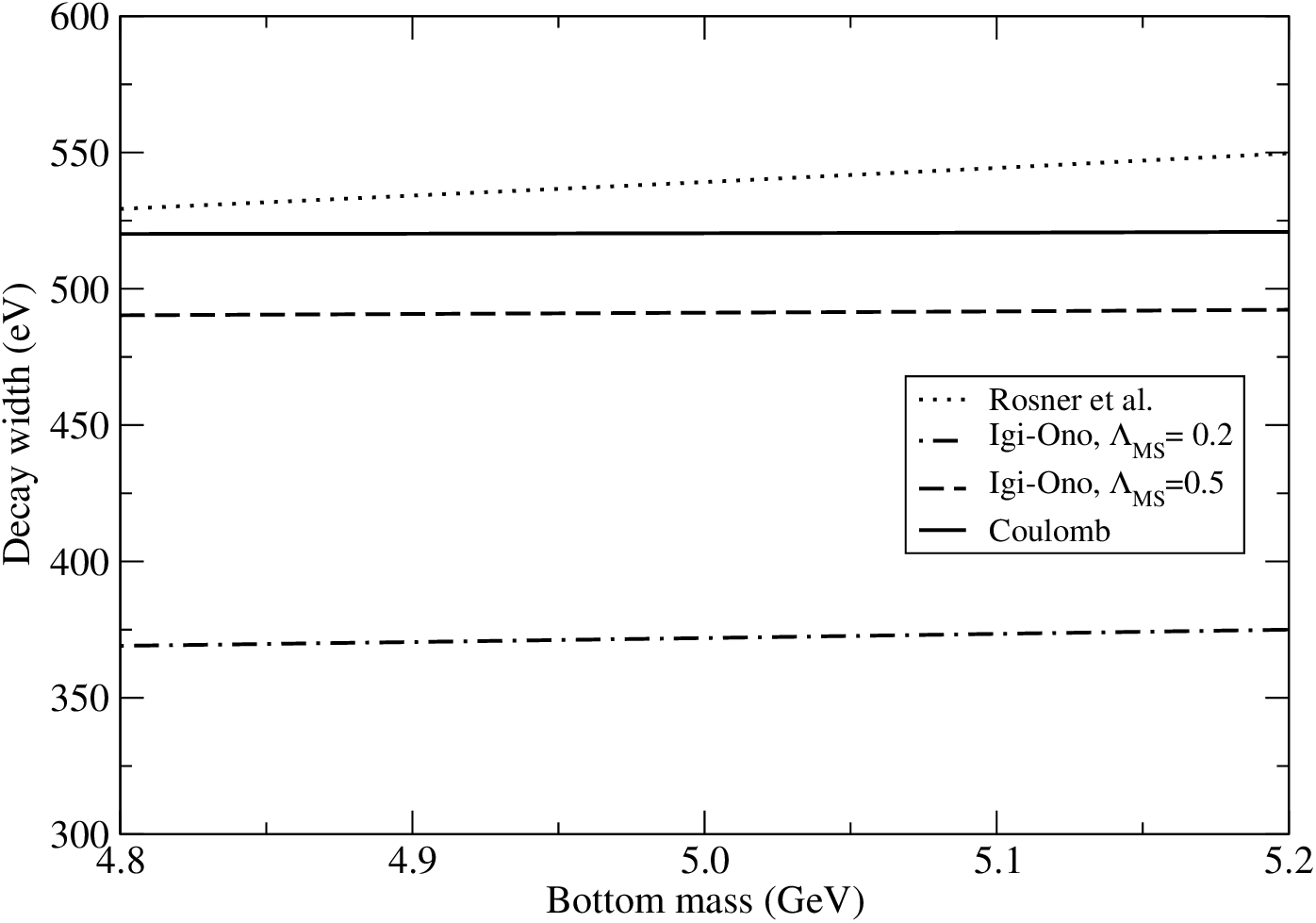}  
\caption{The dependence of $\etab$ decay width to $\gamma\gamma$ in 
\emph{eV} for  different potential models is shown as a 
function of $m_b$~.\label{fig:potentials}}
\end{center}
\end{figure}

We show in fig.~(\ref{fig:potentials}) the predictions for the decay width 
from these  potential models with the correction from 
eq.~(\ref{eq:widsc1l}) at an $\as$ scale $Q=2m_b$~.
For any given model, sources of error in this calculation arise from the
choice of scale in the radiative correction factor and 
the choice of the parameters. Including the
fluctuations of the results given by the different models, we can 
estimate a range of values for the  potential model predictions for 
the radiative decay width  $\Gamma(\etab \to \gamma \gamma)$,
namely

\be
\Gamma(\etab \to \gamma \gamma) =466 \pm 101 \textrm{ eV \mbox{       }.} 
\label{eq:potmodpred}
\ee

\section{Octet component procedure}
We will  present now another approach which admits other components to the 
meson decay beyond the one from the colour singlet picture (Bodwin, 
Braaten and Lepage)~\cite{BBL}. 
NRQCD has been used to separate the short distance scale of 
annihilation from the nonperturbative contributions of long distance scale.
This model has been successfully used to explain the larger than expected 
$\jpsi$ production at the Tevatron and LEP.
According to BBL, in the octet model for quarkonium, the electromagnetic 
and light hadrons ($LH$) decay widths of bottomonium states are given by:

\beqy
\Gamma(\Upsilon \to LH) &=&  
\frac{2\langle \Upsilon | O_1(^3S_1) | \Upsilon \rangle}{m_b^2}  \left (\frac{10 }{243}\pi^2-\frac{10}{27}\right)\as^3 
\times \left [ 1+\left(-9.46\times\frac{4}{3}+ 
\right . \right . \nonumber\\
+12.39-1.161n_f & & \left. \left.  \right) \frac{\as}{\pi} \right]
 +\frac{2\langle \Upsilon | P_1(^3S_1) | \Upsilon \rangle}{m_b^4} \frac{17.32\times \left [20 (\pi^2-9 )\right ]}{486} 
\as^3 
\label{eq:bblups2lh}
\eeqy

\beqy
\Gamma(\Upsilon \to e^+e^-) &=&  
\frac{2\langle \Upsilon | O_1(^3S_1) | \Upsilon \rangle}{m_b^2}  \left [ \frac{\pi}{3}Q^2\alpha^2 \left (1-\frac{13}{3}
\frac{\as}{\pi}\right)\right ] -\nonumber\\
&-&\frac{2\langle \Upsilon | P_1(^3S_1) | \Upsilon \rangle}{m_b^4}\frac{4}{9}\pi Q^2\alpha^2
\label{eq:bblups2ee}
\eeqy

\beqy
\Gamma(\etab \to LH) &=& \frac{2\langle \etab 
| O_1(^1S_0) | \etab \rangle}{m_b^2} 
\frac{2}{9} \pi \as^2 \left [1+\left (\frac{53}{2}-\frac{31}{24}
\pi^2-\frac{8}{9} n_f \right ) \right ] -\nonumber\\
&-&\frac{2\langle \etab | P_1(^1S_0) | \etab \rangle}{m_b^4}\frac{8}{27}\pi
 \as^2
\label{eq:bbleta2lh}
\eeqy

\beqy
\Gamma(\etab \to \gamma \gamma) &=& \frac{2\langle \etab 
| O_1(^1S_0) | \etab \rangle}{m_b^2} \pi Q^4 \aqed^2 
\left [1+\left (\frac{\pi^2-20}{3}\right )\frac{\as}{\pi} \right ]-\nonumber\\
&-& \frac{2\langle \etab | P_1(^1S_0) | \etab \rangle}{m_b^4} 
\frac{4}{3} \pi Q^4\aqed^2
\label{eq:bbleta2gamgam}
\eeqy

There are four unknown long distance coefficients, which can be reduced to
two by means of the vacuum saturation approximation:
\be
G_1 \equiv \langle \Upsilon | O_1(^3S_1) | \Upsilon \rangle = \langle \etab 
| O_1(^1S_0) | \etab \rangle
\ee
\be
F_1 \equiv \langle \Upsilon | P_1(^3S_1) | \Upsilon \rangle = \langle \etab 
| P_1(^1S_0) | \etab \rangle
\ee
correct up to $\mathcal{O}(v^2)$, where ${\vec{v}}$ is  the quark
velocity inside the meson. Since $v$  is of order  $\as(M)$, there is no 
increase of accuracy if the matrix elements are calculated to order
$v^2$ before coefficients are known to order beyond $\as$.

With this position we are able to use the $\Upsilon$ experimental decay 
widths as input in order to determine
the long distance coefficients $G_1$ and $F_1$~. This result in turn
is used to compute the $\etab$ decay widths.

The BBL procedure gives the following decay widths of the $\etab$ meson:

\be
\Gamma(\etab\rightarrow \gamma\gamma)  =364 \pm 8 \pm 13 \textrm{ eV}
\label{eq:bblres2gamma}
\ee
and
\be
\Gamma(\etab\rightarrow LH)  =57.9 \pm 4.6 \pm 2.8 \textrm{ keV ,}
\ee
where the first error comes from the uncertainty on the $\Upsilon$ 
experimental width, the second error from $\alpha_s$~.

The improvement of the error on eq.~(\ref{eq:bblres2gamma})
 with respect to the previous analogous
determination on the $\eta_c$ decay~\cite{NOI2} is due to  better error on the
experimental measures of the $\Upsilon$ decay widths compared to the one 
of the $J/\psi$, and the smaller indetermination on the $\as$ value due to
the higher energy scale involved in the decay. These reasons, together with 
the fact that the potential models used are fitted for the $c\overline{c}$  
system, justifies the improvement of accuracy given in 
eq.~(\ref{eq:bblres2gamma}) compared to the one of 
eq.~(\ref{eq:potmodpred}).

\section{Comparison between models}

\begin{figure}[t]
\begin{center}
\includegraphics*[angle=0,width=0.7\textwidth]{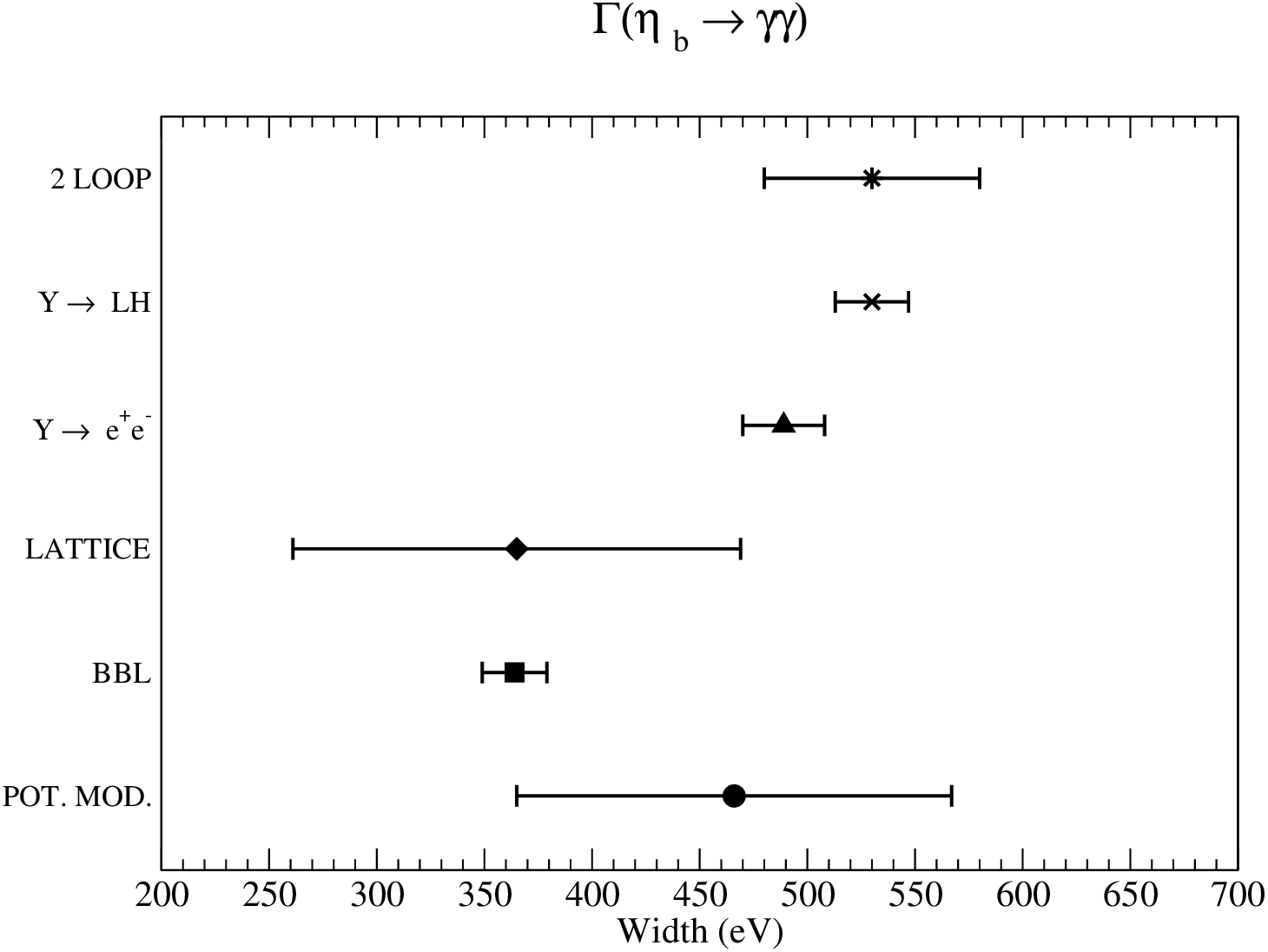}  
\caption{ The $\etab$ two photon width as calculated in this paper 
using (starting from below) Potential Models results, BBL procedure with 
input from $J/\psi$ decay data, Lattice evaluation of
$G_1$ and $F_1$ factors, Singlet picture with $G_1$ obtained from 
$\Upsilon \to e^+e^-$ and $\Upsilon \to LH$ processes 
respectively, and the two--loop enhanced procedure.\label{fig:summary}}
\end{center}
\end{figure}

For comparison we present in fig.~(\ref{fig:summary})
a set of predictions coming from different methods. 
Starting with   potential models, we see that the results are in good
agreement with each other. The advantage of this method is that we are 
giving a prediction from first principles,  without using any 
experimental input. Since there are currently no experimental measures for the
$\etab \to \gamma \gamma$ decay, we shall use this prediction as a 
reference point, as it has proven to be reliable in the case of 
charmonium decay~\cite{NOI2}.
The second evaluation, given by BBL 
using the experimental values of the $\Upsilon$ decay, is on the left limit 
of the potential models value. This is true also for the determination of 
the BBL procedure  with nonperturbative long distance terms taken from
from the lattice calculation~\cite{LATTICE}, affected from a large error. 
The advantage of the latter is that its prediction, like the one from 
potential models, does not make use of any experimental value. 
Next is the point given by the singlet picture from the electromagnetic 
decay of the $\Upsilon$,  aligned with the aforementioned results of the BBL
procedure. 
The point above is obtained also from  the singlet picture with the $\Upsilon$ 
decay into light hadrons, in agreement with the results given from the potential
models. 
We notice that in analogy to the charmonium case (see~\cite{NOI2} and references
therein) the singlet results obtained from the $\Upsilon$ decay are in 
disagreement with each other, in this case by only $1 \sigma$.
The last point from a two--loop enhanced calculation 
given by~\cite{CZARNECKI,ALEPH}
is in agreement with the potential model result and the singlet decay from the
$\Upsilon \to LH$ process.

\section{Conclusions}
The $\Gamma(\etab \to \gamma \gamma)$ decay width prediction of the 
potential models considered gives the value $ 466 \pm 101  \textrm{ eV}$, in
agreement with the naive estimate from the $\Upsilon$ decay given 
by~(\ref{eq:singnaive}). Predictions of the BBL procedure are consistent with
the potential model results, for both the long distance terms $G_1$ and $F_1$ 
extracted from the $\Upsilon$ experimental decay widths and the one evaluated
from lattice calculations. The results from the singlet picture are also
consistent with the potential model results. Finally the two--loop enhanced 
prediction is in good agreement with the potential model results.

\end{document}